\documentclass{JINST}

\usepackage[utf8]{inputenc}
\usepackage{amsmath}

\usepackage[backend=bibtex8,sorting=none]{biblatex}

\renewbibmacro{in:}{}

\title{Physics Performance with the \\CMS Pixel Detector}

\author{Frank Meier, on behalf of the CMS collaboration\\University of Nebraska-Lincoln, Department of Physics \& Astronomy, Jorgensen Hall, 855~N~16th~Street, Lincoln,~NE~68588-0299, USA\\ \email{frank.meier@cern.ch}}

\abstract{A large fraction of the results produced by the LHC experiments during the first run were made possible by precision vertexing detectors. The all-silicon tracking detector of the CMS experiment uses a pixel detector to do vertexing. This conference report uses four physics analyses as examples where pixel detector are used to either reconstruct primary vertices, separate event tracks from pileup, and/or determine displaced vertices.}

\keywords{Particle tracking detectors (Solid-state detectors), Solid state detectors, Performance of High Energy Physics Detectors}

\begin{document}

\section{Introduction}
The first run of the LHC experiments created a wealth of extraordinary results. Silicon-based solid state detectors were used by almost all experiments and were a key to their success. Some used strip topologies, others pixels, in some cases a combination of both popular silicon detector technologies were used, like in the \emph{Compact Muon Solenoid experiment} (CMS). Its pixel detector consists of three barrel layers and two endcaps with two instrumented disks each. The pixels unit cells have a size of $100\,\times150\,\mu\text{m}^2$. A more detailed description can be found in Ref.~\cite{Chatrchyan:2008zzk}.

\section{Physics performance}
Almost all recent physics results by CMS made heavy use of the pixel detector. They either used the detector to a)~reconstruct the primary vertex, b)~associate each track to the vertex of the interesting $pp$-collision (primary vertex) or to the vertices from other $pp$-collisions in the same bunch crossing (``pileup''), or c)~determine displaced (secondary) vertices. In a lot of cases, all three use cases were required for success.

To reconstruct the primary vertices in an event, the tracks from the reconstruction algorithms get analysed for clusters along the longitudinal dimension of the luminous region using a deterministic annealing algorithm, followed by an adaptive vertex fitter to determine the primary vertex position for each cluster. Typical resolutions achieved for primary vertices with at least 50~tracks in samples of \emph{minimum bias} events are 20\,$\mu$m and 25\,$\mu$m in radial and longitudinal direction, respectively~\cite{Chatrchyan:2014fea}. The efficiency to find primary vertices is close to 100\% as soon as more than two tracks are involved. This excellent performance allows physics analyses to rely on accurate positions of the primary vertices and allows for good separation from pileup.

In the following sections, a few examples of physics analyses exploiting the pixel detector are discussed in more detail\footnote{The selection is a personal one, other analyses could have equally well served as examples.}.


\subsection{$B_s\rightarrow \mu\mu$ branching fraction and search for $B^0\rightarrow\mu\mu$}
This analysis focuses on a rare decay of B~mesons, which is especially sensitive to contributions of physics beyond the standard model. Any such contribution would be visible in a significant deviation from the standard model expectation. 
The signal consists of only two muons, required to come from a common secondary vertex, which in turn needs to be separated from the primary vertex because of the lifetime of the $B$-hadron. A multivariate analysis technique is used to select events in that analysis, which is based on \emph{boosted decision trees} (BDT). The first four variables $I$, $I_\mu$, $N_\text{trk}^\text{close}$, and $d^0_\text{ca}$, listed in the excerpt below~\cite{Chatrchyan:2013bka}, measure the isolation of the event in a cone around the $B$-hadron flight direction. This suppresses background events. The $\chi^2$ from the $B$-vertex fit and the distance of closest approach $d_\text{ca}$ of the two muons, are variables that inherit their selective power from the high resolution of the detector.
\begin{quote}
For each BDT, a number of variables is considered and only those found to be effective are included. Each of the following 12 variables, shown to be independent of pileup, are used in at least one of the BDTs: $I$; $I_\mu$; $N_\text{trk}^\text{close}$; $d^0_\text{ca}$; $p_T^{\mu\mu}$; $\eta_{\mu\mu}$; the $B$-vertex fit $\chi^2$ per degree of freedom (dof); the $d_\text{ca}$ between the two muon tracks; the 3D pointing angle $\alpha_\text{3D}$; the 3D flight length significance $\ell_\text{3D}/\sigma(\ell_\text{3D})$; the 3D impact parameter $\delta_\text{3D}$ of the $B$ candidate; and its significance $\delta_\text{3D}/\sigma(\delta_\text{3D})$, where $\sigma(\delta_\text{3D})$ is the uncertainty on $\delta_\text{3D}$. The last four variables are computed with respect to the primary vertex. Good agreement between data and MC simulation is observed for these variables. \dots
\end{quote}
A distinct property of pixel detectors is their capability for vertexing in 3D. The power of this comes from being able to distinguish true vertices of the two muons from vertices that are unreal but would have been taken as such in a 2D only measurement -- because in a 2D projection, any pair of sufficiently close oppositely charged tracks form at least one perfect vertex, but only in 3D real vertices can be identified as such. The almost square shape of the CMS pixel unit cells offer comparable resolution in both the $r\phi$ plane and the $z$ direction, a property being very helpful for this analysis.


\subsection{Higgs boson properties in $H\rightarrow ZZ\rightarrow 4\ell$}
The decay of a Higgs boson in four leptons is an example of a very clean signature at the LHC, with all the Higgs daughters originating from one vertex. Different combinations of lepton flavors are considered: $4e$, $4\mu$ or $2e2\mu$. It is thinkable to make this analysis in the $4\mu$ channel only, by using just a detector with muon chambers but no specialised vertexing detector. The reduced resolution in such an approach would be a limiting factor for success, as this would make it hard to suppress background events from in-flight decays of hadrons. The following excerpt shows how this analysis makes use of the 3D vertexing capabilities of the pixel detector~\cite{Chatrchyan:2013mxa}:

\begin{quote}
In order to suppress leptons originating from in-flight decays of hadrons and muons from cosmic rays, all leptons are required to come from the same primary vertex. This is achieved by requiring $\text{SIP}_\text{3D}<4$ where $\text{SIP}_\text{3D}\equiv\text{IP}_\text{3D}/\sigma_{\text{IP}_\text{3D}}$ is the ratio of the impact parameter of the lepton track ($\text{IP}_\text{3D}$) in three dimensions (3D), with respect to the chosen primary vertex position, and its uncertainty.
\end{quote}
The result shows a nice, narrow resonance at a mass of $125.6 \pm 0.4 (\text{stat}) \pm 0.2 (\text{syst})$\,GeV with a width of $\leq 3.4$\,GeV (95\% confidence limit). Moreover, the analysis was able to determine the production cross section and spin-parity properties, which were found to be compatible with standard model predictions.

\subsection{VZ production cross section in $VZ\rightarrow Vb\bar{b}$}
The precise knowledge of standard model processes is very important for any search of new physics, rare decays or channels with large background contributions. One such case is the $VZ$ production cross section, where $V$ is either a $W$ or a $Z$ vector boson~\cite{Chatrchyan:2014aqa}. Its knowledge is a key ingredient towards the measurement of $H\rightarrow b\bar{b}$.

A key ingredient here is the ability to use reliable $b$-tagging, a domain of pixel detectors. As an example for the performance, the efficiency for tagging a $b$-jet has been observed to be about $80-85\%$ with a misidentification rate of 10\%. Using tighter parameters the efficiency drops to about $45-55\%$ at the benefit of a misidentification rate of 0.1\%~\cite{Chatrchyan:2012jua}.

The increase of the LHC beam energy will change the typical transverse momentum of such jets to higher values, which calls for even better detectors in the future. Higher momentum jets are more collimated, resulting in more frequent merging of charge deposits by different tracks in a single cluster. New and continuously improved algorithms help but finer pitched sensors will clearly be a better choice. In addition, the trend to higher $p_T$ and to innermost pixel layers at ever smaller radii challenges the reconstruction algorithms, as the likelihood for $b$-jets to fly past the first layer increases. Many current $b$-jet tagging algorithm work under the assumption that there must be a signal in the first layer, which may no longer be the case for a significant fraction of $b$-jets.

\subsection{$\Lambda_b$ lifetime}
The high performance of the CMS pixel detector is also useful in precision measurements. The $\Lambda_b$ lifetime has long been measured to be too short compared to theoretical predictions. CMS measured the lifetime of this baryon in the channel $\Lambda_b\rightarrow J/\psi \Lambda$, where its products decay as $J/\psi\rightarrow\mu\mu$ and $\Lambda\rightarrow p\pi$~\cite{Chatrchyan:2013sxa}. Three vertices take part in this measurement: the primary vertex to measure the begin of the life of a particular $\Lambda_b$, the secondary vertex of the $J/\psi$ to determine its decay, and the $\Lambda$ decay vertex with its characteristic long lifetime ($c\tau\approx 7.89$\,cm) to select events and suppress background. The momentum information of all four tracks is required to determine the decay time in the particle's rest frame. To extract all required observables, kinematic vertex fitting using 3D information has been used. The result showed a lifetime compatible with theoretical expectations, in contradiction to earlier measurements but in line with more recent results.

\section{Final remarks and conclusions}
This selection of analyses acted as examples of the performance of the CMS pixel detector. This is the result of a well-crafted design, translated into working hardware, operated, properly calibrated and aligned by a dedicated team of specialists. Without the visionary people courageous enough to start with this development years ago, these physics results would not have been possible. Pixel detectors have been shown to work reliable in a high pileup environment, with increasing track multiplicities and in an increasingly harsh radiation environment.


\printbibliography

\end{document}